\documentstyle[a4,12pt,amssymb,epsfig]{article}

\newcommand{\e}{\mbox{e}} 
\newcommand{\Pcobe}{{\cal{P}}^{1/2}_{\cal{R}}}
\def\fun#1#2{\lower3.6pt\vbox{\baselineskip0pt\lineskip.9pt
        \ialign{$\mathsurround=0pt#1\hfill##\hfil$\crcr#2\crcr\sim\crcr}}}

\renewcommand\({\left(}
\renewcommand\){\right)}
\renewcommand\[{\left[}
\renewcommand\]{\right]}

\newcommand\eq[1]{Eq.~(\ref{#1})}
\newcommand\eqs[2]{Eqs.~(\ref{#1}) and (\ref{#2})}

\newcommand\ee{\end{equation}}
\newcommand\be{\begin{equation}}
\newcommand\eea{\end{eqnarray}}
\newcommand\bea{\begin{eqnarray}}

\newcommand\GeV{\,\mbox{GeV}}
\newcommand\MeV{\,\mbox{MeV}}

\newcommand\mpl{M_{\rm P}}

\newcommand\Mpl{M_{\rm P}}

\newcommand{\lsim}{\mbox{\raisebox{-.9ex}{~$\stackrel{\mbox{$<$}}{\sim}$~}}}

\newcommand{\gsim}{\mbox{\raisebox{-.9ex}{~$\stackrel{\mbox{$>$}}{\sim}$~}}}

\def\dslash{\not{\hbox{\kern-2pt $\partial$}}}
\def\Dslash{\not{\hbox{\kern-4pt $D$}}}
\def\Oslash{\not{\hbox{\kern-4pt $O$}}}
\def\Qslash{\not{\hbox{\kern-4pt $Q$}}}
\def\pslash{\not{\hbox{\kern-2.3pt $p$}}}
\def\kslash{\not{\hbox{\kern-2.3pt $k$}}}
\def\qslash{\not{\hbox{\kern-2.3pt $q$}}}

 \newtoks\slashfraction
 \slashfraction={.13}
 \def\slash#1{\setbox0\hbox{$ #1 $}
 \setbox0\hbox to \the\slashfraction\wd0{\hss \box0}/\box0 }

\def\ee{\end{equation}}
\def\be{\begin{equation}}

\def\calf{{\cal F}}

\def\calp{{\cal P}}
\def\calr{{\cal R}}
\def\calpr{{\calp_\calr}}

\newcommand\sub[1]{_{\rm #1}}

\newcommand{\ie}{{\it{i.e. }}}
\newcommand{\eg}{{\it{e.g. }}}
\newcommand{\meta}{{|\eta_0|}}

\renewcommand{\(}{\left(} \renewcommand{\)}{\right)}
\renewcommand{\[}{\left[} \renewcommand{\]}{\right]}
\renewcommand{\thefootnote}{\fnsymbol{footnote}}

\begin{document}
\thispagestyle{empty}
\vspace*{.5cm}
\noindent
\hfill {\tt hep-ph-ph/0502047}\\
\vspace*{1cm}
\begin{center}
{\Large\bf Hilltop Inflation}\\[.8cm] {\large Lotfi
Boubekeur and David H. Lyth}\\[.6cm]
{\it Physics Department, Lancaster University, Lancaster LA1 4YB,  U.K.}
\end{center}
\vskip 1 cm

\begin{abstract}

We study `hilltop' inflation, in which  inflation takes place  near a
maximum of the potential.  Viewed as a model of inflation after the
observable Universe leaves the horizon (observable inflation)  hilltop
inflation is rather generic.  If the potential steepens
monotonically, observable hilltop inflation gives a tiny  tensor
fraction ($r\lsim 0.002$).  The usual $F$- and $D$-term models may
easily be transmuted to hilltop models by Planck-suppressed terms,
making them more natural.  The only commonly-considered  model of
observable inflation  which is definitely {\em not} hilltop is 
tree-level hybrid inflation. Viewed instead as an initial condition, 
we explain  that  hilltop inflation is more generic than seems to 
have been previously recognized, adding thereby to the
credibility of the  idea that eternal inflation provides the
pre-inflationary initial condition.

\end{abstract}


\renewcommand{\thefootnote}{\arabic{footnote}}

\section{Introduction}
\label{s1}

It is now clear that the Universe is spatially flat, and that it 
possesses an almost scale-invariant adiabatic density perturbation before
cosmological scales come inside the horizon. Both of these features
suggest that the initial condition for the Universe is set during some era
of almost-exponential inflation, beginning around the time that the 
observable Universe leaves the horizon and ending several tens of
$e$-folds later. We shall call this era {\em observable} inflation,
to distinguish it from inflation that may have occurred earlier.

Slow-roll inflation provides an attractive mechanism for inflation. 
According to this paradigm, the energy density during inflation is 
dominated by the inflaton potential $V(\phi)$. 
The trajectory in field space is parameterised by the canonically 
normalized inflaton field $\phi$, which rolls sufficiently slowly 
that $V$ is almost constant. This generally requires the potential 
to be flat to high degree of accuracy, which is difficult to achieve 
in the context of particle physics motivated models \cite{treview,book}. 
Not only the $\eta$-problem prevents inflation from occurring in 
supergravity models, it would also produce a spectrum far from scale 
invariant. A possible solution to this tension is to liberate the inflaton 
from the task of producing the observed density perturbations. Indeed, 
density perturbations can originate from the quantum fluctuations of 
any scalar field which is effectively massless during inflation. In the 
simplest case (inflaton paradigm), it is the inflaton field itself that 
will deliver the totality of curvature perturbation. However, in general 
there could be other light fields which could contribute to, or even 
dominate, the density perturbation as well. These alternative scenarios 
for density perturbation have distinctive features in the CMB that will 
be probed with the advent of the next generation of experiments like 
Planck \cite{planck}.

Coming back to the flatness conditions, an obvious strategy to satisfy the 
flatness conditions is inflation occurring near a (local) maximum of the 
potential. We shall call this situation `hilltop' inflation. This kind of 
model has been proposed after discovering the graceful exit problem of old  
inflation and was dubbed ``new inflation'' \cite{newinflation}. The first 
advantage of such a scenario is that the slow-roll conditions can be 
satisfied much more easily. While the first slow-roll condition follows 
automatically from the fact that inflation starts out from a local maximum, 
the second slow-roll condition can be relaxed \cite{ks1} to $|\eta|< 6$, 
leading to what has been called fast-roll inflation \cite{fastroll}. This 
allows inflation to occur with natural values of $\eta$, closer to what one 
expects from a generic supergravity theory, alleviating thus the 
$\eta$-problem.

Besides easing the slow-roll conditions, these models seem to be
favoured  both theoretically and experimentally. On the experimental
side, hilltop  inflation is part of what is called ``small-field
models'', a class of single  field inflationary models characterized
by the negative curvature of their potential. The  latest analysis
\cite{data} seems to favour them among the other types (large  field
and hybrid models). On the theoretical side, owing to the fact that
the variation of the inflaton field during inflation is  $\lesssim
\mpl$, their potential  is much more understandable in terms of
effective field theory. Furthermore,  potentials of this type occur in
particle physics models where symmetries are  broken either
explicitly or spontaneously. Finally, hilltop inflation naturally 
incorporates eternal inflation making the discussion of 
initial conditions, which is an issue in most inflation models, 
completely irrelevant. 

The purpose of this paper is to study this class of models and their
experimental signatures. We will focus mostly on hilltop inflation as
a model of {\em observable} inflation. After recalling the slow-roll
formalism and its observational constraints, we show that observable
hilltop inflation is unlikely to  give a significant  tensor
perturbation.  We then examine in detail the case of modular
inflation, characterised by a VEV  $\phi\sim\mpl$.  Then we  examine
Natural Inflation, which seems to be the only well-motivated model
which might give a VEV $\phi\gg \mpl$.  Next, we show how
Planck-suppressed terms may easily convert  the usual $F$- and
$D$-term models to hilltop, removing much of their fine-tuning. Then
we point out that running-mass inflation provides  another hilltop
model.  We end with  some remarks about hilltop inflation viewed as an
initial condition for non-hilltop models of observable inflation,
followed by a summary of our findings.

\section{Slow-roll Inflation}
\label{s2}

Let us recall the slow-roll formalism. In the space of the scalar 
fields there could be either a unique inflationary trajectory, or 
family of possible trajectories.  In the latter case we focus on 
the trajectory chosen by Nature (in, at least, our part of the universe).
The inflaton field $\phi$ measures the distance in field space along this
trajectory. It may or may not be convenient to define the origin of $\phi$
as a fixed point of the symmetries, corresponding to an extremum of the
potential $V(\phi)$.

Slow-roll inflation is defined by the flatness conditions 
\footnote{As usual $\mpl\equiv(8\pi G_N)^{-1/2}= 2.4 \times 10^{18}$ GeV.}

\be
\epsilon \equiv \frac{M_P^2}{2} \(V'(\phi)\over V(\phi)\)^2 \ll 1, \quad 
{\textrm{and}}\quad |\eta|\equiv M_P^2 \left| V''(\phi)\over V(\phi) \right| \ll 1
\label{flatness}
\,,
\ee
together with the slow-roll approximation $3H\dot\phi\simeq - V'(\phi)$ 
and the critical density condition $\rho =3H^2\mpl^2$.  
From these it follows that $3H^2\mpl^2 \simeq V(\phi)$. 

Slow-roll inflation ends when either one of the flatness conditions 
\eq{flatness} is violated, or else the potential is de-stabilized in 
the direction of some  `waterfall' field (hybrid inflation). 

The value of $\phi$ when a scale, defined by the wavenumber $k$, leaves 
the horizon is given by 
\be 
N(k)  = \mpl^{-1} \int^\phi_{\phi\sub{end}} 
\frac{d\phi}{\sqrt{2\epsilon(\phi)}} 
\label{nexp} \,.
\ee 
Here $N(k)$ is the number of $e$-folds, from horizon exit to the end of
inflation at $\phi=\phi\sub{end}$, which satisfies $dN\simeq -d\ln k$.
Cosmological scales leave the horizon during about $10$ $e$-folds,
starting with the exit of the whole observable Universe which corresponds
to $k=H_0$ (the present value of the Hubble parameter). We are interested
then in $N\equiv N(H_0)$. For a typical post-inflationary cosmology, the 
required number of $e$-folds is \footnote{It is understood that all  
quantities appearing in the right hand side of the equations like the 
slow-roll parameters, the Hubble rate and the height of the inflationary 
potential, should be evaluated at the epoch when the relevant scale leaves 
the horizon, though its slow variation during inflation is not usually 
significant.}
\be
N\simeq  60- \log \(\frac{10^{16}\GeV}{V^\frac14}\)
\,.
\ee

We shall use this estimate of $N$ except where stated. The differential form 
of \eq{nexp} is
\be 
dN =  \mpl^{-1}
\frac{d\phi}{\sqrt {2 \epsilon} }
\,.
\ee
To keep the field theory under control,  inflation models are typically
constructed so that the variation of $\phi$ is exponentially small on the 
Planck scale, at least while cosmological scales leave the horizon.
Then  $\epsilon$ is also exponentially small. 

Around the time of horizon exit, the vacuum fluctuation of the inflaton
field is converted to a classical perturbation, which is practically
Gaussian with spectrum $\calp_\phi = (H/2\pi)^2$. After horizon exit this 
corresponds to a position-dependent shift back and forth along the inflaton
trajectory, which corresponds to a time-independent curvature perturbation, 
with spectrum
\be
\calp_\calr(k)    =
\frac1{24\pi^2\Mpl^4}\frac{V}{\epsilon} 
\label{ss5} 
\,.
\ee
The spectral index of this perturbation is
\be
n_S-1 \equiv  {d\log \calp_\calr \over d\log k}=  2\eta - 6\epsilon 
\label{ss6}
\,,
\ee
There may be also a tensor perturbation, whose spectrum is some fraction 
$r$ of $\calpr$ is given by
\be
r=16\, \epsilon
\label{r}
\ee
\section{Observational bounds}

The observed adiabatic density perturbation is equivalent to a spatial
curvature perturbation $\calr$, whose spectrum $\calpr$ and spectral
index $n_S$ are determined by observation. According to  a
recent  analysis \cite{data}, at $1-\sigma$,
\bea
\Pcobe &\simeq&  5 \times 10^{-5}
\,,
\label{cobe}\\
-0.048 & < &  n_S-1  < 0.016
\,,
\label{observ}
\eea

The Planck satellite will give an accuracy $\Delta n_S\simeq 
0.01$ by the end of the decade \cite{planck}. 

On (large) cosmological scales, data \cite{data} gives for the tensor 
perturbations
\be
r< 0.47 \qquad\qquad \textrm{95$\%$  c.l.}
\label{tensors}
\ee
The Planck satellite will give only $r<0.1$ or so \cite{planck}, but projects 
dedicated to detecting the tensor should give better than $r<0.01$ by the end 
of the decade
\cite{Taylor:2004hh}, the ultimate limit \cite{ultimate} being more like 
$r<10^{-4}$. Instead of a limit on $r$ there could of course be a detection, 
but we focused on the limit because that will be the eventual outcome of 
observation according to a wide class of inflation models 
\cite{notensor,treview,book}. 

\paragraph{Inflaton Paradigm} 
According to the standard assumption, the inflaton contribution given by 
\eq{ss5} is solely responsible for the observed  curvature perturbation.
In that case \eqs{ss5}{ss6} provide a powerful constraint on the shape and
magnitude of the inflationary potential;
\bea
(V/\epsilon)^{1/4} &=& 6.6\times 10^{16}\GeV 
\label{cmbnorm}\\
-0.048 &<&  2\eta - 6\epsilon  < 0.016 \label{indexbound}
\,.
\eea
For the typical case that $\epsilon$ is negligible,  {\em the spectral index 
measures the curvature (second derivative) of the inflationary potential}. 
There is also another bound on the height of the potential which can be 
obtained by combining the bound on tensors \eq{tensors} and \eq{cmbnorm}
\be
V^{1/4}<2.71 \times 10^{16} \GeV
\label{gw}
\ee
The present precision of data does not permit to distinguish between the 
different shapes of inflaton potentials. However the data favours the 
small-field ($\epsilon\sim0,$ $\eta<0$) over the large-field  (\eg 
Linde's chaotic model \cite{chaotic}) and the hybrid models.
\paragraph{Non-Inflaton Paradigms}
The inflaton paradigm  is inevitable if the inflationary trajectory
is unique, since in that case the potential in orthogonal directions
will be too steep for the vacuum fluctuation to be converted into a classical
perturbation. If instead there is a family of trajectories, the vacuum
fluctuation in each direction orthogonal to the inflaton 
is converted at horizon exit into a Gaussian 
classical perturbation, with the same spectrum $(H/2\pi)^2$ as the
inflaton perturbation. Such a perturbation does not contribute to the 
curvature perturbation at horizon exit, but it {\em may} generate 
a curvature perturbation later which gives a significant (even dominant) 
contribution to the total observed curvature perturbation.
This may happen in various ways. It may happen during inflation. 
For that to  occur, the  orthogonal field must significantly affect the 
inflationary dynamics corresponding to `two-field' or `two-component' 
inflation  where the trajectories are curved \cite{treview,book}. 
Alternatively, it may happen during preheating \cite{preheating}, or 
during a reheating process (modulated decay \cite{decay}) or during 
the run-up to some  reheating {\em not} caused by the
inflaton decay (the  curvaton mechanism \cite{curvaton}).

If the observed curvature perturbation receives a contribution from one or 
more of the orthogonal fields, \eq{cmbnorm} becomes only an  upper bound;
\be
(V/\epsilon)^{1/4} < 6.6\times 10^{16}\GeV 
\label{cmbbound}
\ee
while \eq{indexbound} is  weakened. If an orthogonal 
contribution is  dominant, \eq{indexbound} does not apply at all, being 
replaced by (Eqt.~(115) in \cite{treview}) 
\be
n_S-1=2\eta_{\sigma\sigma} -2\epsilon
\label{ncurv}
\,,
\ee 
where $\eta_{\sigma\sigma}\equiv \mpl^2\,\partial^2 V/\partial\sigma^2$.
Then  {\em the spectral index has nothing to do with the curvature of the
inflaton potential}. We can also derive an upper bound on tensors using 
\eq{cmbbound} 
\be
r < 16\epsilon
\label{rbound}
\,,
\ee
the equality being attained if the curvature perturbation is generated
by the inflaton and we are back to \eq{r}. If instead the inflaton 
contribution is negligible, $r$ is negligible and the tensor fraction 
is unobservable.

\section{Hilltop inflation: Generalities}
Hilltop inflation is supposed to take place near a maximum of the 
potential, which means that the potential will have the form 
\footnote{See \eg \cite{Linde:1984cd} for an early model il the 
context of $N=1$ supergravity.}
\bea
V(\phi) = V_0 -\frac12 m^2 \phi^2 + \cdots = 
V_0 \( 1 - {1\over 2} |\eta_0| \( \frac{\phi}{\mpl}\) ^2 +\cdots \)
\,,
\label{vhill}
\eea
with $V\simeq V_0$ and the dots indicating higher order terms in the
power series expansion. The maximum $\phi=0$ is not necessarily a
fixed point of  internal symmetries and has been chosen as the origin
only for convenience.  The tachyonic mass $m$ is characterised by
$\eta_0<0$, which is the value of  $\eta$ at the maximum.  For the
most part we focus on the case that the mass term dominates, at least
while cosmological scales leave the horizon.  Even if it  does not
dominate there is no reason to expect the contribution $\eta_0\subset
\eta$ to be strongly cancelled by the contribution of the  additional
terms in \eq{vhill}.  Barring such a cancellation, slow-roll inflation
requires $\meta\ll 1$,  which is in mild conflict with the value
$\meta\sim1$ expected in a generic supergravity theory
\cite{sugramass,sugramassafter}.

Before moving on to specific models, we present a bound on the tensor
fraction $r$, which is valid for any hilltop model in which the 
slope parameter $\epsilon(\phi)$ increases monotonically \footnote{This is 
not the case for Natural Inflation.}. 
In such a model \eq{nexp} gives when cosmological scales leave the horizon
\be
2\epsilon < \frac1{N^2} \( \frac{\phi\sub{end}}{\mpl} \)^2
\,,
\ee
which by using \eq{r} corresponds to  
\be
r<0.002 \( \frac{60}{N} \)^2 \( \frac{\phi\sub{end}}{\mpl} \)^2
\label{rhillbound}
\,. 
\ee
In terms of the height of the hill this bound translates to
\be
V^{1/4} < 7.0 \times 10^{15}\GeV \( \frac{60}{N} \)^{1/2}
\( \frac{\phi\sub{end}}{\mpl} \)^{1/2}
\label{vhillbound}
\,.
\ee
If instead we specialise to a quadratic hilltop potential of the 
type \eq{vhill}, we can derive a much more stringent bound using \eq{r} 
\be
r<0.0003 \(60 \over N\)^2 \(\phi\sub{end}\over\mpl\)^2.
\label{rb}
\ee
Neither of the factors in brackets in \eq{rb} and \eq{vhillbound} is likely 
to be much bigger than 1, therefore we conclude that hilltop inflation is 
unlikely to give a detectable tensor fraction. 

\section{Modular hilltop inflation}
\label{s3}
We will consider various possibilities, beginning in this section
with the case that inflation ends at $\phi\sub{end}\sim\mpl$. We have in
mind particularly the case of modular inflation \cite{bgmodular,banksmodular},
 in which $\phi$ is a light string modulus having a potential of the form 
$\Lambda^4 {\cal F}(\phi/\mpl)$, the {\em typical} values of $\cal F$ and its 
derivatives being of order 1. Inflation with this sort of potential can be 
achieved if ${\cal F}^\prime$ and ${\cal F}''$ are anomalously small for some 
range of $\phi$. We are considering hilltop inflation which ensures the first 
condition, but ${\cal F}''$  needs to be suppressed because its generic value 
would correspond to $\meta\sim 1$.

Consider first the extreme case that the  mass term dominates until 
$\phi\sub{end}\sim \mpl$.
This gives 
\be
\phi_N \sim \mpl e^{-N|\eta_0|}
\,.
\ee
If the inflaton generates the curvature perturbation,
\bea
\calp_\calr^{1/2} &=& 
\frac
1{2\sqrt3\pi} {V_0^{1/2}\over \mpl^2 |\eta_0| e^{-N\meta}} \\
n_S-1 &\simeq & -2|\eta_0| 
\,.
\eea
Inserting the observational value of $\calp_\calr$ gives $V_0(\meta)$, which 
is plotted in Figure ({\ref{final}). The  observational constraint on $n_S$
requires $|\eta_0|< 0.024$. 
We have repeated this calculation using the much more accurate fast-roll 
approximation \cite{supernatural,fastroll,Lyth-Stewart}. It gives
\begin{eqnarray}
\phi(N)&\simeq&  
\phi\sub{end}\,\frac{\delta+1}{2 \delta}\; \e^{-3N(\delta-1)/2},\\
\delta&\equiv&\sqrt{1-\frac43\eta_0} \\
\Pcobe(k) &\simeq& 2^\Delta \;\frac{\Gamma(3/2 + \Delta)}{\Gamma(3/2)} \;\frac{H}{2\pi \Delta \;\phi} \;\left(\frac{k}{a\,H}\right)^{-\Delta}\,,\label{p}\\
\Delta&\equiv& {3\over 2} (\delta -1), 
\end{eqnarray}

Setting again $\Pcobe$ equal to the observational value gives
the solid curve in Figure (\ref{final}). 

Let us make some comments on Figure (\ref{final}) . The bound
\eq{rbound}  on the tensor fraction is almost saturated, corresponding
to the fact that inflation can take place with 
$\phi\sub{end}\sim\mpl$ with practically constant  $\epsilon$.
The constraint on $\meta$ coming from the observed spectral index is
tight,  and will become tighter if observation continues to push the
lower bound on  $n_S-1$ towards zero. As we have seen though, this
constraint disappears if  the curvature perturbation is generated by
some field orthogonal to the  inflaton. Then, the only constraint on
$\meta$ is that the potential is big  enough.  An absolute bound is
$V_0^{1/4}> 10\MeV$ is demanded by  nucleosynthesis, corresponding to
the bottom horizontal line. This allows  $|\eta_0|$ up to $10.54$ or
so. Most likely though, the inflation scale is  higher than the scale
of supersymmetry breaking in the vacuum, corresponding to say
$V_0^{1/4}\sim 10^{10}\GeV$ allows $|\eta_0|\sim 1$. This conclusion 
is somehow strengthened by the fact that in the naive curvaton scenario 
the height of the potential cannot be less than $10^{12}-10^{13}$ \GeV 
\cite{lowcurv}. 

In any case, the liberation \cite{dl} of hilltop inflation from the
requirement  that the inflaton generates the curvature perturbation is
seen to  dramatically increase the allowed parameter space. Finally,
we note  that the slow-roll approximation is quite adequate in the
favoured regime $V_0^{1/4}\gsim 10^{10}\GeV$ and gives in any case a
fairly good approximation.

\begin{figure}[t]
\centering
\includegraphics[scale=.8]{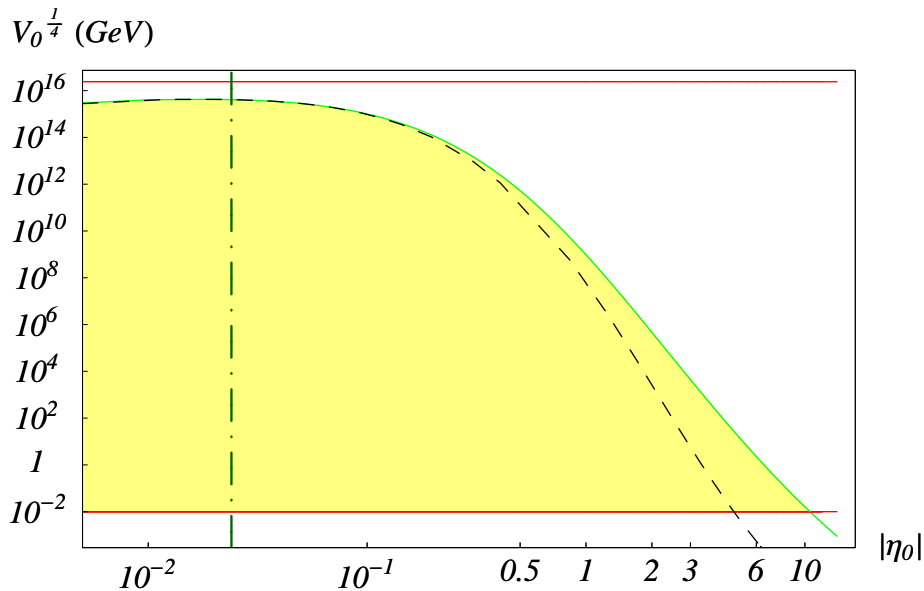}
\caption{\footnotesize The upper bound on the height of hilltop inflation. 
The dashed (black) curve is obtained using the slow-roll approximation 
while the continuous (green) curve is obtained using the better fast-roll 
approximation. The top horizontal line represents the bound from the absence 
of a tensor signal in the CMB, which is seen to be automatic for hilltop 
inflation. The bottom horizontal line represents the absolute lower bound 
coming from BBN. If the inflaton perturbation generates the observed curvature 
perturbation, the observational bound on the spectral index places 
$|\eta_0|$ to the left of the vertical dash-dot (dark green) line.}
\label{final}
\end{figure}

All of this assumes that the mass term dominates until the end of inflation.
Now  we suppose  instead  that 
the mass term dominates  only while cosmological
scales leave the horizon, the potential thereafter becoming steeper.
Remarkably, the observational constraint in this much more general case 
is just the same as in the previous case. This is
because, from \eq{nexp}, the steepening will reduce the value of $\phi$ at
horizon exit, which will lower $V_0$ so that the curve in Figure (\ref{final}) 
is still an upper bound on the potential. Also, if the inflaton generates the
curvature perturbation, $\meta$ must still lie to the left of the vertical
line. For future reference we note that all of this remains true in the
case $\phi\sub{end}<\mpl$; for the moment though we are focusing on
modular inflation corresponding to $\phi\sub{end}\sim\mpl$.

We can consider also the case that a higher power of $\phi$ dominates the
mass term when cosmological scales leave the horizon. If there is no
symmetry $\phi\to-\phi$, the power will presumably be 3. This case has
been considered in \cite{ars}, where $\phi$ is supposed to be a modulus
corresponding to \footnote{Since the origin in this case is an inflexion 
point instead of a maximum, we will focus on the region $\phi>0$.}
\be
V = V_0 \( 1 - \lambda_3 \phi^3/\mpl^3 + \cdots \)
\,,
\ee
with $\lambda_3\sim 1$. If the inflaton generates the curvature
perturbation this gives $V_0^{1/4}\sim 3\times 10^{14}\GeV$, and spectral
index $1-n_S= 4/N\simeq 0.07$ which violates the $1-\sigma$ bound
(\ref{observ}). (If the curvature perturbation is generated by an
orthogonal field, the only constraint is $V_0^{1/4}< 3\times 10^{14}\GeV$.)

If there is a  symmetry $\phi\to-\phi$, the power will presumably be 4,
corresponding to
\be
V = V_0 - \lambda\phi^4 + \cdots
\,.
\ee
If the inflaton generates the curvature perturbation, $\lambda\simeq
10^{-12}$ and $1-n_S = 3/N\gsim 0.05$, which is allowed by the $1-\sigma$
bound (\ref{observ}). It seems not to have been noticed before that such a
coupling can be quite natural for a modulus, corresponding to
\be
V=V_0 \( 1 -\lambda_4 \phi^4/\mpl^4 + \cdots \)
\,,
\ee
with $\lambda_4\sim 1$. Indeed this gives $\lambda \sim V_0/\mpl^4$
which has the required value for $V_0^{1/4}\simeq 10^{15}\GeV$.
(If the curvature perturbation is generated an orthogonal field
the only constraint is $\lambda<10^{-12}$, corresponding to
$V_0^{1/4}<2\times 10^{15}\GeV$.) 
Notice that for both the cubic and quartic
potential the tensor fraction is negligible, which actually
is the case for any power \cite{notensor}. 

We shall illustrate the applicability of this discussion with the two most
recent and detailed examples of modular inflation. Both of them are
two-component models, because a complex modulus is considered and the
trajectory in field space is curved.  Consider first the model of
\cite{andreirace}, inspired by KKLT stabilization \cite{kklt}. The
inflaton perturbation is assumed to dominate the curvature perturbation,
which may be a reasonable approximation since  the inflationary
trajectory does not seem to be very strongly curved. The inflationary
potential depends on several parameters, whose values are fine-tuned to
make $\meta\ll 1$. The authors consider as an example a particular set of
parameters corresponding to $\meta=0.015$, and an initial condition giving
altogether $137$ $e$-folds of inflation. With this choice the inflationary
trajectory is beginning to steepen significantly by the time that
cosmological scales leave the horizon, giving $1-n_S\simeq 0.05$ (instead
of $1-n_S=0.03$ which would be the case if there were no steepening) and
$V_0^{1/4}\simeq 10^{14}\GeV$ (instead of $V_0^{1/4}\simeq 10^{16}\GeV$).
The value of $V_0$  corresponds more or less to the string scale, which has
been chosen by hand to fit the observed curvature perturbation. This model 
as it stands has the following two problems. First, fine-tuning
is required to make $\meta$ small enough to satisfy the spectral index
constraint. Second, $V_0^{1/4}$ is much bigger than the value
$V_0^{1/4}\sim 10^{10}\GeV$ which might be expected if the height of the
potential is related to the supersymmetry breaking scale of the MSSM
(see however \cite{andreinewrace}). We would like to point out that
both of these problems are caused by
the assumption that the inflaton field perturbation generates the
curvature perturbation, which need not be the case.

The other model that we want to mention \cite{ks1,ks2} works in the 
context of supergravity, valid more or less up to the Planck scale,
and assumes gravity-mediated supersymmetry breaking corresponding to 
$V_0^{1/4}\sim 10^{10}\GeV$. The tree-level  mass-squared 
is supposed to have the generic supergravity value  corresponding 
to $\meta\sim 1$, with the origin a fixed point the symmetries. The 
interactions generate a loop correction which turns the maximum into a 
crater, whose rim corresponds to a maximum in each radial direction. 
Since the mass only runs logarithmically, the maximum typically corresponds 
to $\meta\ll 1$, satisfying the slow-roll inflation requirement
without fine-tuning. In calculating the curvature perturbation both 
components are taken into account. For a typical trajectory the 
contribution of the orthogonal component would not be much
bigger than that of the inflaton, which for the desired normalization
$V_0^{1/4}$ would not generate a big enough curvature perturbation.
Instead, the  trajectory is chosen to have strong curvature (justified 
{\em a postieri} by considering the volume of inflated space),
so that the curvature perturbation is generated almost entirely
by the orthogonal component, allowing the desired low scale
$10^{10}\GeV$. Since the curvature perturbation comes from the 
orthogonal component, the spectral index is given by
\eq{ncurv}, and depending on the choice of parameters it may or may not be
indistinguishable from  1.\footnote {To be precise, the spectral index is 
given by a modified form \cite{ss} of \eq{ncurv} which takes into account 
the non-canonical normalization.}

\section{Natural/chaotic inflation}
\label{s4}

In this section we will consider the case where hilltop inflation ends 
only at $\phi\gg\mpl$. In that case a generic effective field theory is 
not under control because the potential generically receives contributions 
$\lambda_n\phi^{n+4}/{\mpl^n}$ which all matter at $\phi\gsim\mpl$.
Only one theoretically-motivated mechanism has been proposed 
for dealing with them \cite{accr} where $\phi$ is a PNGB with a periodic 
potential and a scale of spontaneous breaking \footnote{It is claimed 
\cite{banks03} that $f\gg \mpl$ is unlikely in the context of string 
theory, at least if $\phi$ is a modulus. See however \cite{Kim:2004rp}.} 
$f\gg \mpl$. This model corresponds to what has been called Natural 
Inflation \cite{natural}.

The periodic potential is
\be
V= \frac12 V_0 \( 1 + \cos(\sqrt{2\meta} \phi/\mpl) \)
\,, 
\ee
which in the small angle limit reduces to \eq{vhill}. The slow-roll 
parameters are then 
\bea
\epsilon &=&  \frac1{2N} \frac{2N\meta}{e^{2N\meta}-1} \label{naturaleps}\\
\eta &=&  \epsilon -  \meta 
\,,
\eea
where $\phi$ at horizon exit is given by 
\be
\sin\( \sqrt{\eta_0\over 2} {\phi\over\mpl}\)= \sqrt{1 \over 1 + \eta_0} \,e^{-N \eta_0}
\ee

Let us consider the observational constraints on Natural Inflation,
on the assumption that the inflaton generates the density perturbation. 
The spectral index is 
\be 
n_S  = 1 -4\epsilon - 2\meta
\,.
\ee

In Figure (\ref{naturaln}) we show $n_S-1$ against $\meta$.
(We set $N=60$ for simplicity since this model cannot give
$V^{1/4}$ far below $10^{16}\GeV$ if the inflaton generates
the curvature perturbation.)

In the regime  $N\meta\ll 1$, inflation takes place near the minimum of the 
potential, corresponding to the `chaotic inflation' potential. In the 
opposite case it takes place near the maximum, corresponding to hilltop 
inflation. The present observational bound corresponds to 
$\eta_0\lesssim 0.02$, or $N\meta\lesssim 1.2$. This means 
that observable inflation takes place nearer to the maximum than 
the minimum of the potential. Fixing $N=60$, we can get the corresponding 
threshold value $\eta_0=.0057$ that distinguish between the two regimes, 
which is within the sensitivity of Planck. Thus Natural Inflation can 
generate chaotic inflation, and as a matter of fact it is so far the only 
theoretically-motivated proposals doing that. 
From \eq{naturaleps}, this bound on $\meta$ means that Natural Inflation 
predicts a tensor perturbation which will definitely be observable in the 
future. The model may though turn out to be indistinguishable from chaotic 
inflation.
 
\begin{figure}[t]
\includegraphics[scale=.8]{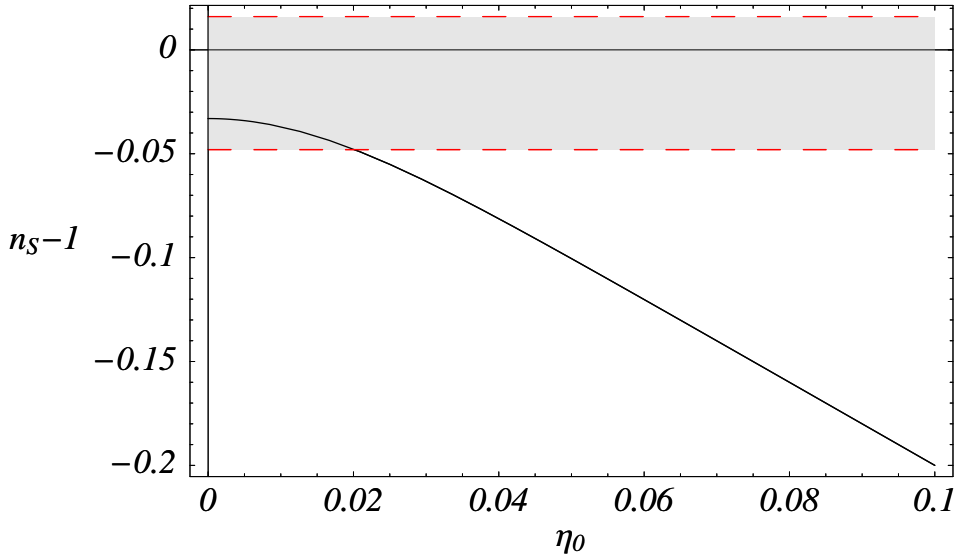}
\caption{\footnotesize $n_S - 1$ versus $\eta_0$ for the Natural Inflation 
scenario. The shaded region corresponds to the present WMAP/SDSS bound 
\cite{data}.}
\label{naturaln}
\end{figure}

We emphasise that all of these conclusions assume that 
the curvature perturbation is generated by the inflaton. 
If Natural Inflation is liberated from this requirement
it is constrained only by the bound \eq{cmbbound} on the height of the 
potential.

\section{Converting  $F$- and $D$-term models to hilltop}
\label{s5}
\begin{figure}[t]
\centering
\includegraphics[scale=.8]{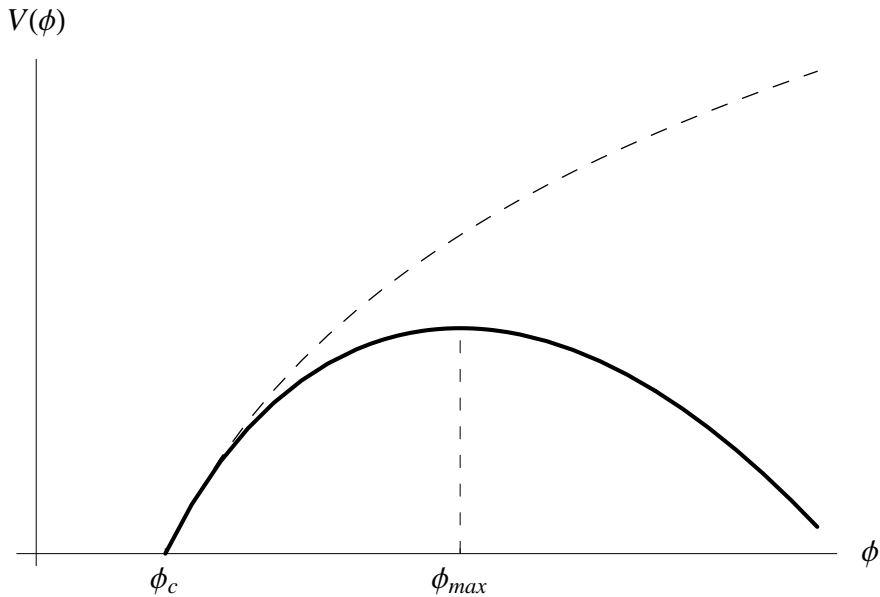}
\caption { \footnotesize Sketch of the inflationary potential for the 
$F$/$D$ - term scenario when including the non renormalizable terms 
(continuous line) and the original potential (dashed line). }
\label{logsketch}
\end{figure}
In a class of inflation models \cite{treview,book} the slope of the potential
levels out without actually turning over, being of the form
$V'\propto 1/\phi^p$ or $V'\propto \exp(-p\phi)$. This class of covers a 
wide range of particle physics motivated models\cite{treview,book}. To see 
how things can work, it will be enough to focus on supersymmetric hybrid 
`$F$-term' and `$D$-term' inflation, which are perhaps the best-motivated 
case. 

\paragraph{$F$-term scenario} We begin by considering $F$-term inflation.
The original model \cite{dss} is based on the following superpotential and 
minimal Kahler potential 
\cite{cllsw}
\bea
W(S,\,\psi,\,\bar\psi)&=&\lambda S(\psi\bar{\psi}-\Lambda^2)\,,
\label{superpot}\\
K(S,\,\psi,\,\bar\psi)&=&|S|^2+|\psi|^2+|\bar\psi|^2\,.
\eea
The waterfall field $\psi$ might be a GUT Higgs, whose VEV $\Lambda
\simeq 10^{16}\GeV$ defines the GUT scale.
The  canonically normalised inflaton field is $\phi\equiv \sqrt2|S|$, 
and for $\phi>\phi\sub c \simeq \sqrt{2}\Lambda$ there is inflation with a 
perfectly flat  tree-level potential \cite{cllsw}.
The  loop correction gives 
\begin{equation}
V(\phi)=V_0 \left[ 1+ \frac{\lambda^2}{8\pi^2} \log\left(
\frac{\phi}{Q}
\right)
\right]
\,,
\label{vlog1}
\end{equation}
with 
  $V_0=\lambda^2 \Lambda^4$. Here 
$Q$ is the renormalization scale which should be  choosen to 
 make the {\em magnitude} of the loop correction small while cosmological
scales leave the horizon. The derivatives of $V$ are independent of $Q$.
If $\lambda \lsim 4\pi\Lambda/\mpl$  slow-roll inflation continues until 
$\phi=\phi\sub c$. We focus on the case 
$\lambda \gsim 4\pi\Lambda/\mpl$. 
In that case slow-roll inflation ends at the 
value of $\phi$ given by
\be
\phi\sub{end} \sim  \frac{\lambda}{2\pi\sqrt 2}\, \mpl
\,,
\ee
corresponding to $|\eta|\sim 1$. Inflation continues for a while, as
 $\phi$ oscillates about zero, but this lasts only until $\phi\sim \phi\sub c$
which takes a negligible number of $e$-folds
(about  $\ln (\phi\sub{end}/\phi\sub c)$). Then the waterfall field 
ends inflation in the usual way.
Cosmological scales leave the horizon when 
\be
\phi\simeq \sqrt{N\lambda^2\over 4\pi^2} \, \mpl
\label{vv}
\,.
\ee

If the inflaton is responsible for the curvature perturbation,
 Eq. (\ref{cmbnorm}) and (\ref{vv}) give the  CMB normalization 
\begin{equation}
\Lambda\simeq \( \frac{50}{N}\)^{1/4} \times 6 \times 10^{15} \GeV
\label{dss}
\,.
\end{equation}
This is independent of the coupling $\lambda$, and taking the uncertainties
into account it justifies
the identification of $\Lambda$ with the GUT scale.
The spectral index is given by 
\be
1-n_S={1\over N} \gsim 0.02
\,.
\ee

If some other field is responsible for the curvature, $\Lambda$ is lower
and the observed spectral index gives no constraint on the potential.

The model as we described it so far  assumes the superpotential \eq{superpot},
and the canonical Kahler potential.
 Generically  one expects that the superpotential contains higher powers of 
$\phi$, and that the  Kahler function  contains  non-canonical  terms.
These terms are completely out of control at  $\phi\gsim \mpl$.
To avoid this, we shall assume initially $\phi\ll\mpl$. In view of \eq{vv}, 
this  requires  
\be
\lambda\ll 1  
\label{lambound}
\,.
\ee

With $\phi\ll\mpl$,  powers of $\phi$ can be forbidden by
a $Z^n$  $R$-symmetry with suitably high $n$. (In  the context of 
string theory this is more reasonable than imposing a continuous 
$U(1)$ $R$-symmetry which would forbid all powers.)  What about the 
 non-canonical terms in the 
Kahler potential? With $\phi\ll\mpl$, their effect will be to just
generate a mass-squared $m^2$ with generic magnitude of order
  $V_0/\mpl^2$.
Including it the potential becomes
\begin{equation}
V(\phi)=\lambda^2 \Lambda^2 \left[ 
1+ \frac{\lambda^2}{8\pi^2} \log\left(
\frac{\phi}{Q}
\right) + {1\over2 M_P^2}\eta_m \phi^2
\right]\,,
\label{vlog}
\end{equation}
with $\eta_m\equiv m^2\mpl^2/V_0$ generically of order 1
in absolute value magnitude.
The case of  positive $\eta_m$ has been investigated already \cite{panag}.
Here we look for the first time at the case of
 negative $\eta_m$, which corresponds to hilltop inflation
as in Figure (\ref{logsketch}). The
maximum of the potential is at 
\be
\phi\sub{max} ={ \lambda\over 2\pi\sqrt{2 |\eta_m|}}\mpl
\label{phim}
\,.
\ee

In order to keep $\phi\sub{max}\ll\mpl$ we need 
\be
|\eta_m|\gg \frac{\lambda}{2\pi\sqrt 2}
\label{ss} .
\ee

Near the maximum the potential  has the form \eq{vhill} (after
shifting the origin of $\phi$) with $\eta_0=2\eta_m$.  Inflation is
supposed to take place while $\phi$ rolls from the  maximum to
smaller values, and the log term  steepens the potential  at
$\phi\sim\phi\sub{end}$ which ends slow-roll inflation as described
earlier. For inflation to occur near the maximum we need $\meta\ll
1$. This means  that the mass has to be somewhat below the generic value.

Let us assume first that the inflaton is responsible for the curvature 
perturbation. Then the CMB normalisation determines the VEV $\Lambda$;
\be
\Pcobe
=\sqrt{4N \over 3} \left(\frac {\Lambda}{M_P}\right)^2\, f(N\,\eta_m),
\label{pmodfterm}
\ee
where $f(x)$ is given by
\be
f(x)\equiv \left(\frac{\e^{-2x} - \e^{-4x}}{2x}\right)^{1/2}
\,.
\ee
This curve is plotted in Figure (\ref{hybrid}). 
The spectral index is
\bea
1-n_S &=& \meta  \( 1 + \frac1{1-\exp(-N\meta)}\) 
\,,
\eea
where $\eta_0 = 2 \eta_m$ is the value of $\eta$ at the hilltop. In
the regime $\meta\gsim 1/N \simeq 0.02$ we have $1-n_S\simeq 2\meta$,
corresponding to the situation that cosmological scales leave the
horizon before the potential steepens appreciably.\footnote {In
accordance with our earlier discussion, the potential $V_0^{1/4}$  in
this regime must be  below, which is seen to be the case by virtue of
$V_0^{1/4} <\Lambda$.}  The observational constraint is therefore
$2\meta < 0.048$, which is indicated by a vertical line.

Now suppose instead that some other field is responsible for the
curvature  perturbation. The curve in Figure (\ref{hybrid}) becomes
an upper bound on  $\Lambda$,  and the spectral index gives no
constraint.  The allowed range is still small if we identify $\Lambda$
with the GUT scale. On the other hand, this  identification need not
be compulsory. In particular, we may choose $\Lambda$ to be the scale
of Peccei-Quinn symmetry breaking,  which is in the range from
$10^{10}\GeV$ (indicated by the horizontal line), up to around
$10^{13}\GeV$ (or  even up to $10^{15}\GeV$ if final reheating is
below $1\GeV$). Thus $\meta$ need not be so fine-tuned.

\begin{figure}[t]
\centering
\includegraphics[scale=.8]{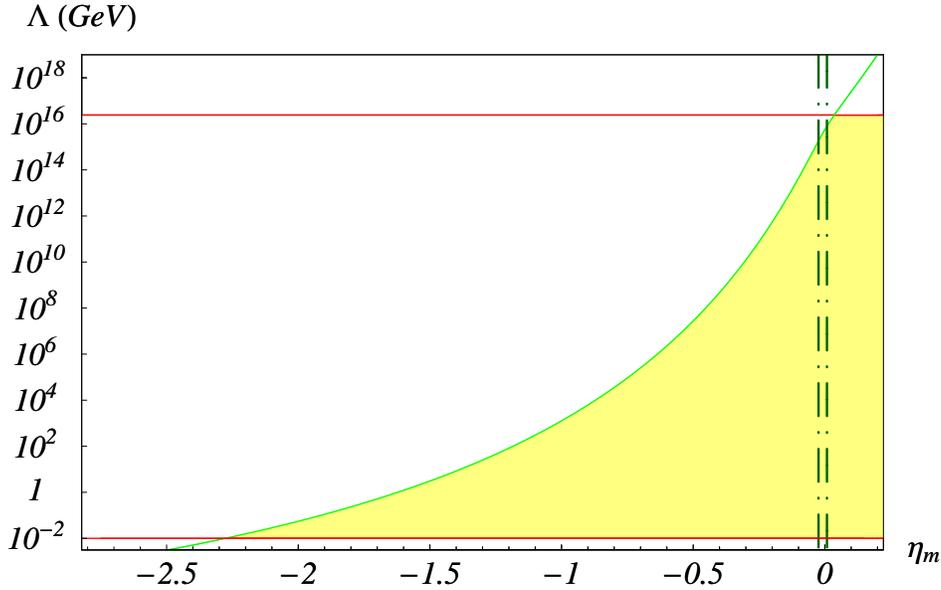}
\caption{\footnotesize 
The VEV of the waterfall field  as a function of $\eta_m$. Any point 
in the light yellow area is allowed. The upper horizontal (red) line 
stands for the bound on tensor fluctuations while the lower one stands 
for the BBN bound. The vertical dot-dash lines are for the the bound on 
$n_S$.}
\label{hybrid}
\end{figure}
So far we have required $\phi\ll\mpl$.
Now we consider instead the case
$\phi\sim\mpl$. Then, 
non-canonical terms will generate a potential
\begin{equation}
V(\phi)=\lambda^2 \Lambda^2 \left[ 
1+ \frac{\lambda^2}{8\pi^2} \log\left(
\frac{\phi}{Q}
\right) + {\cal F}(\phi/\mpl)
\right]\,,
\label{vlog2}
\end{equation}
where $\calf (x)$ is a function with value and derivatives typically of order 
1 at $x\sim 1$.

This potential has not usually been taken seriously in the 
literature, where instead at most a quartic additional term has
been considered \cite{panag}, with positive sign (and  magnitude
corresponding to the minimal Kahler potential).
In truth one should take on board the fact that $\cal F$ is unknown.
Precisely because it is unknown, {\em it is equally reasonable to suppose 
that $\cal F$ has a negative slope, sufficient to generate a maximum of the 
potential at $\phi\sim\mpl$}. Then we will have hilltop inflation,
with the potential steepened by the log term. The results we
have obtained will still be more or less correct; up to a numerical 
factor of order unity (which depends on the unknown  form of $\cal F$) 
the parameter $|\eta_m|$ can be identified with the parameter $\meta$ 
defining the curvature at the hilltop.

\paragraph{$D$-term scenario} We end by considering the case of 
$D$-term inflation \cite{ewandtof,dterm} which is based on the 
superpotential $W(S,\,\psi,\,\bar\psi)=\lambda S \psi\bar{\psi}$ 
and canonical Kahler. The resulting potential is also of the form 
(\ref{vlog1}), with the following replacements: 
$\lambda=g/\sqrt{2}$, where $g$ is a $U(1)$ gauge coupling and 
$\Lambda$ is the Fayet-Iliopoulos term $\sqrt{\xi}$. In this case 
however $\cal F$ can be generated from the gauge kinetic function 
\cite{mydterm}  as well as any non-negligible $F$-term \cite{lrdterm}. 
Since the gauge coupling is presumably not small, one definitely
expects $\phi\sim\mpl$ for the $D$-term case, which means that 
the correction $\cal F$ should not be ignored. On the assumption
that $\cal F$ has positive slope, this has been regarded as a serious 
problem for $D$-term inflation  \cite{mydterm,km}. In contrast, the negative
slope that we have investigated here removes this particular problem
for $D$-term inflation. (For a thorough review of the rather delicate
status of $D$-term inflation, see \cite{dreview}.)

\section{Small-field hilltop inflation}
\label{s6}
  The discussion so far has focused on models with $\phi\sub{end}$
at most an order of magnitude or so below $\mpl$. We end our discussion
of hilltop inflation by considering models with $\phi\sub{end}$ several
orders of magnitude below $\mpl$. In such models, 
Planck-suppressed terms are usually completely negligible, even 
though the potential is very flat \cite{treview}.

Almost all models of this type so far proposed are of the hybrid type,
where the potential at the end of inflation falls in the direction of some
waterfall field different from the inflaton. A non-hybrid model of this 
type is described in \cite{dr} (see \cite{dl} for a fuller discussion).
In the  original hybrid inflation model \cite{hybrid} the potential is
\be
V = V_0 - {1\over2} m_\chi^2 + {1\over4}\lambda\chi^4 + {1\over2}m^2\phi^2 + 
{1\over2}\lambda' \chi^2\phi^2
\label{hybfull}
\,,
\ee
where $\chi$ is the waterfall field and both $m^2$ and $\lambda'$ 
(also $m_\chi^2$ and $\lambda$) are supposed to be positive. This 
leads during inflation to
\be
V=V_0 +\frac12m^2\phi^2 
= V_0 \( 1 + \frac12 \eta_0 \frac{\phi^2}{\mpl^2} \) \label{hybinf}
\,,
\ee
where $\eta_0>0$ is the constant value of $\eta$. (When we later
include additional terms it will be the value of $\eta$ at the minimum.)
This is not hilltop inflation. One way of converting it to hilltop 
\cite{inverted} is to reverse the signs of $m^2$ and $\lambda'$. This 
possibility has been realised in the model of \cite{steve1}, but in general 
one expects positive $\lambda$ which closes off this route to hilltop 
inflation. It might therefore appear that hilltop inflation is exceptional 
in the context of hybrid inflation.

In fact that is not the case, because the tree-level potential
typically  receives a significant loop contribution. One  such
contribution  certainly comes from the coupling of the inflaton to the
waterfall field, and at 1-loop this contribution   has been shown
\cite{myhybrid} to dominate the tree-level term in a large region of
parameter space. In a significant part  of that region, the
contribution is so big that it forbids inflation altogether. It has
further been pointed out \cite{supernatural} that owing to the
flatness of the potential the 2-loop correction may considerably
extend this forbidden region.

We conclude that \eq{hybinf} should not be regarded as the generic
hybrid inflation potential, because it is quite likely to be modified
by a loop correction which may or may not come from the coupling to
the waterfall field.  The form of the correction depends on how
supersymmetry is broken.  If spontaneously broken global supersymmetry
is a good approximation we get the potential of the last section. If
instead softly broken global supersymmetry is a good approximation,
$m^2$ is converted to  $m^2(\phi)$, depending logarithmically on
$\phi$. If $m^2$ goes through  zero at some point, there will
typically be a nearby  maximum or a minimum, near which $\eta$ can be
small enough for  inflation to occur even if the tree-level value
(identified with $m^2(\mpl)$) corresponds to the generic supergravity
value $|\eta|\sim 1$. This is the running mass model
\cite{ewanrunning}. The potential is of the form
\be
V(\phi)= V_0 \[ 1 + \frac12\eta_0
 \( \ln\frac{\phi}{\phi_0} - \frac12 \)
\]
\,,
\label{run}
\ee
where $\eta_0$ is the position $\phi_0$ of the maximum or minimum.\footnote
{In the notation of \cite{ourrunning2}, $\eta_0=c$.}
A maximum, corresponding to hilltop inflation,
 is preferred theoretically because a minimum tends to require a fine-tuned
end to inflation \cite{ewanrunning,ourrunning2}. 

This model requires $\meta \gsim 0.1$ corresponding to significant running
of the mass, otherwise the potential will not acquire the desired 
maximum. One might think that observation rules out such a value,
but that is not yet the case \cite{ourrunning2} because
the  potential has  a point of inflexion a bit to the left of the maximum.
If observable scales leave the horizon while the field is passing 
through this point, the spectral index passes through zero
(corresponding to a minimum in the spectrum), and its
scale-dependence can be in accordance with observation.

\section{Eternal hilltop inflation}
\label{s7}
We have seen that hilltop inflation can be thought as a model of 
{\em observable} inflation, beginning when our observable Universe 
leaves the horizon. An alternative might be that the inflaton is 
initially near a hilltop, but has moved far away by the time that 
the observable Universe leaves the horizon with the result that 
the potential for observable inflation is not at all of the 
hilltop form. 

To study this interesting possibility, consider first the
tree-level hybrid potential  defined by  \eqs{hybfull}{hybinf}.
Assume that   the loop correction is  negligible, and that the 
parameters are such that inflation ends at   $\phi\sub c$
some  orders of magnitude below $\mpl$, and that the slow-roll
condition $\eta_0\ll 1$ is very well satisfied. 
Then our Universe leaves the horizon at $\phi_*=\phi\sub\exp(N\eta_0)$ 
which is still some orders of magnitude below $\mpl$.

As $\phi$ increases beyond $\phi_*$, higher-order terms in the potential
will eventually become significant. Assuming that the quartic term
$\lambda\phi^4$ is small enough (to be precise, that $\lambda\lsim 
V_0/\mpl^4$) one may expect that these terms become important only at the 
Planck scale so that the potential has the form
\be
\label{hill}
V(\phi) = V_0 \( 1 + \frac12 \eta_0 \frac{\phi^2}{\mpl^2}
+ \calf{(\phi/\mpl)} \)
\,
\ee
with $\calf(x)$ and its derivatives of order one in the regime
$x\sim 1$.

Analogously with the discussion at the end of Section \ref{s5},
the slope $\calf'$ might be either positive or negative.
If it is positive one may expect that $V$ increases monotonically.
In that case, the potential {\em may} support inflation all the way up to
the Planck scale $V\sim \mpl^4$. The condition for that to be so is
that the flatness conditions \eq{flatness} are satisfied. 
Because we are in the regime $\phi\gg \mpl$ the flatness conditions 
are satisfied for quite generic potentials but they are not at all inevitable.
For instance, they are satisfied by $V\propto\phi^p$ for all $p>0$ 
(monomial or `chaotic' inflation), but they are satisfied by 
$\phi\propto \exp(\sqrt{2/p}\phi/\mpl)$ only for $p>1$.

In this case there {\em may}  be a regime where the first flatness condition 
is very well satisfied \footnote{For $V\propto\phi^p$ this regime clearly 
exists, while this is not the case for $V\propto \exp(\sqrt{2/p}\phi/\mpl)$.}, 
to be precise with  $\epsilon \lsim (V/\mpl)^{1/2}$. In such a regime the 
quantum fluctuation experienced by any  sub-horizon sized region overcomes the 
classical slow-roll, giving presumably an indefinitely large inflating region.
This is the phenomenon of eternal inflation \cite{eternal}. 
If there is a regime of the potential permitting eternal inflation, all 
discussion about the probability of its actual occurrence 
becomes irrelevant, because the infinite inflating volume outweighs any
finite improbability for the process to start. 

Whether inflation at $\phi\gsim\mpl$ happens in this first case (let
alone eternal inflation)  depends on the form of the potential.  The
only guidance on this score comes from string theory, in the case that
$\phi$ is a modulus. At least in that case, inflation with a
monotonically increasing potential does not seem likely, the potential
of the canonically-normalized modulus being typically an exponential
of an exponential \cite{polchinskibook}.

If inflation does not occur at $\phi\gsim\mpl$, it may be very
improbable  for inflation to start at all, since in order to achieve
that one has to  create the observable Universe with energy density
far below the Planck scale. (As noted in \cite{Linde:2004nz} such a
conclusion  is not inevitable if the universe is periodic.)  At the
opposite extreme, if {\em eternal} inflation occurs, then as we noted
already probability considerations become irrelevant.
\begin{figure}[t]
\centering \includegraphics[scale=.8]{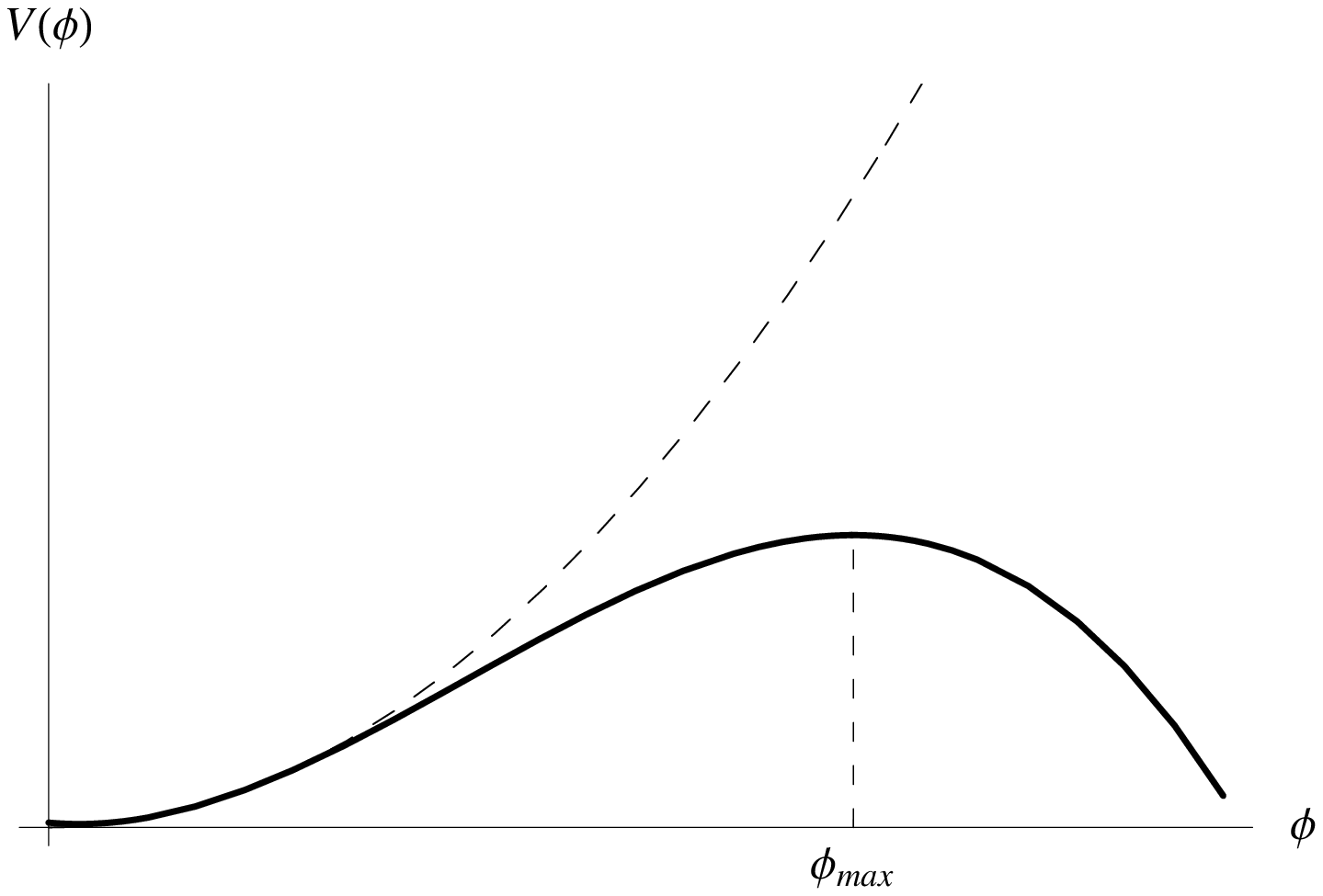}
\caption{\footnotesize Sketch of the hilltop potential of \eq{hill}
when  $\calf'$ is negative.}
\label{treesketch}
\end{figure}

The second case, which does not seem to have been  explicitly
considered before, is that $\calf'$ is negative,  generating a maximum
of the potential.  This case,  illustrated in Figure
(\ref{treesketch}), is quite analogous with the case we considered
earlier (potential \eq{vlog2} and (Figure (\ref{logsketch})) except
that descent from the hilltop happens  long {\em before} the
observable Universe leaves the horizon. In contrast  with the first
case, inflation in this second case is likely, simply because   the
generic value of $\eta$ at the hilltop is  $\meta\sim 1$. It is not
necessary to postulate any special behaviour of the function $\calf$,
beyond  the behaviour $\calf'\sim -1$ postulated for its first
derivative at  $\phi\sim\mpl$. In the case of string theory moduli
such behaviour is not  at all unlikely. Even more interesting is the
fact that inflation in this  case inevitably starts out as  eternal
inflation, provided only that there  {\em exist}  regions of space
where  $\phi$ is initially  sufficiently close  to  the hilltop. Even
if the quantum fluctuations were absent,  eternal inflation would
occur in patches where $\phi$ is extremely close to  the the hilltop
because the volume generated by the regions with smallest  $\phi$
would always outweigh the volume generated by regions with larger
$\phi$ \cite{topinf}. We emphasize again that all discussion about the
probability for our Universe to be located within such a region is
irrelevant, owing to the indefinitely large volume created by eternal
inflation. Altough the above considerations were made about 
non-observable inflation, they apply equally to observable hilltop inflation.

The above discussion leads to the following conclusions, for the
particular  tree-level model that we have considered. First, initial
hilltop inflation  looks about as likely as the alternative of a
monotonically increasing  potential. Second, hilltop inflation is more
desirable, because it  practically obviates the need to consider the
initial probability  for our Universe to inflate.

Before ending this section, we comment on models which cannot be converted 
to hilltop.  Surveying the range of potentials that have been proposed
for observable inflation, as described for instance in
\cite{treview,book}, one can see that we have now considered most of
them, but not quite all.  We have yet to consider tree-level hybrid
inflation with a higher power \ie   $V = V_0 \( 1 + c\phi^p \)$, with
$p$ an integer bigger than $2$, as well as  the potential $V = V_0 \(
1 +c/\phi^p \)$, with positive $p$ which may  correspond to dynamical
symmetry breaking. The constant $c$ is positive for  both
potentials. In contrast with the (quadratic) tree-level model which we
considered earlier, these models cannot easily be modified to  give
initial hilltop inflation. For the second model this is because the
motion is away from the origin, and in the first model it is because
the  flatness conditions fail before $\phi$ gets to the Planck scale.
Because of these features, it is not even the case that observable
inflation can follow on smoothly from inflation with $\phi\gg
\mpl$. It seems fair to say therefore that these models are
disfavoured compared with all of the others.  Such a conclusion is
perhaps not unwelcome, because they  are also distinguished from the
others by the fact that the predicted spectral index (for the case
that the inflaton gives  generates the curvature perturbation) is not
determined by the form of  the potential. Instead it is typically
given by 
\be 
n_S-1 = \frac{p-1}{p-2} \frac2{N\sub{total} - N }\,, 
\ee
where $N\sub{total}$ is the total number of $e$-folds of inflation
which in  these models is finite. Practically any $n_S>1$ is therefore
allowed.  Thus, the situation of these models in relation to
observation is rather unhappy compared with that of the others,
making it perhaps welcome that they seem also to be {\em a priori}
less likely.

\section{Conclusion}
\label{s8}

Although, slow-roll is a dominant paradigm for inflation and the 
generation of density perturbations, it is difficult to achieve in 
the context of particle physics models.  We have shown that inflation 
starting from a local maximum of the potential is able to address 
two of the most serious problems of slow-roll inflation model building: 
fine tunning and initial conditions. The first problem is addressed 
by liberating the inflaton from the task of generating density 
perturbations, allowing thus more natural values of $\eta$. The second 
problem is addressed through eternal inflation, which occurs naturally 
in these models and  makes the discussion about initial probabilities 
irrelevant. We have shown that most models are easily converted 
to hilltop. In particular, we have illustrated how that works is 
supersymmetric hybrid inflation using Planck-suppressed terms. We 
studied also models which are already of the hilltop type like Natural 
and modular inflation. The only exception are models with potentials 
$V\propto (1 + c\phi^p)$, with $c>0$ and $p$ integer $p>2$ or $p<0$ that 
cannot be converted to hilltop. We have also derived observational constraints 
on the amount of gravitational waves produced which will be unlikely 
to be detected by the next generation of experiments, except for Natural 
Inflation.
We conclude that hilltop inflation, occurring  while cosmological
leave the horizon or only much earlier, is both generic and desirable.

\subsection*{Acknowledgments}
We acknowledge Andrei Linde for useful comments, and Andrew Liddle for
drawing our \mbox{attention} to  \cite{Taylor:2004hh}. We are supported 
by PPARC grants PPA/G/O/2002/00469, PPA/V/S/2003/00104,
PPA/G/O/2002/00098 and PPA/S/2002/00272.

{\footnotesize

\end{document}